# Computational Design of Boron-Free Triangular Molecules with Inverted Singlet-Triplet Energy Gap

M. W. Duszka, M.F. Rode, and A.L. Sobolewski*
*Institute of Physics, Polish Academy of Sciences, Warsaw, Poland*

**Abstract**

A novel, computationally designed, class of triangular-shape organic molecules with an inverted singlet-triplet (IST) energy gap is investigated with the aid of *ab initio* methods of electronic structure theory. The considered molecular systems have a form of cyclic oligomers and their common feature is electronic conjugation localized along the molecular rim. Analysis of vertical transition energies from the electronic ground state, as well as from the lowest excited singlet and triplet states of selected molecules, is conducted. The results underscore the significance of optimizing excited-state geometries in theoretical models to accurately describe the optoelectronic properties of the IST molecules, particularly in relation to their applications in OLEDs.

**Graphical TOC**

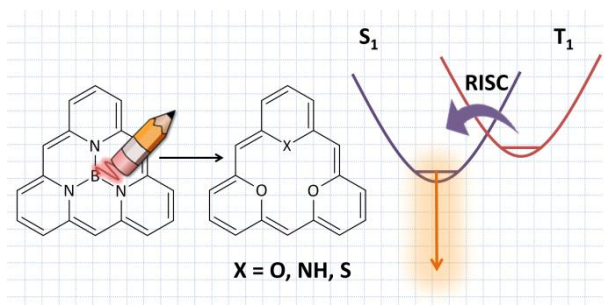

Chemical modifications of the triangular boron-carbon-nitride (**BCN**) molecules that remove the insulating BN interior but conserve electronic conjugation at the molecular rim keep the inverted singlet-triplet nature of the systems.

* corresponding author, e-mail: sobola@ifpan.edu.pl

**Introduction**

Within the domain of molecular physics, Hund's multiplicity rule, stating that triplet excited states should exhibit lower energy levels than their singlet counterparts with identical orbital configuration, has long served as a guiding principle. This fundamental principle has been a reliable rule governing the ordering of excited states in organic molecules and it was assumed that violations of Hund's multiplicity rule in the excited states of organic compounds are exceptionally rare. Recent developments, however, have confirmed the existence of energy inversion of the first singlet ($S_1$) and triplet ($T_1$) excited states, known as inverted singlet-triplet (IST) states, in a number of stable closed-shell organic molecules.[1,2]

The significance of organic molecules featuring nearly degenerate or inverted $S_1$ and $T_1$ states extends well beyond theoretical chemistry, being of relevance for the development of chromophores for organic light-emitting diodes (OLEDs). Many organic chromophores currently under examination for OLED applications exhibit small positive $S_1$-$T_1$ energy gaps,[3–6] relying on reverse intersystem crossing (RISC) from $T_1$ to $S_1$, resulting in thermally activated delayed fluorescence (TADF) at room temperature.[7,8] The emergence of organic IST chromophores may open the door to a new generation of OLED devices, capitalizing on the potentially intense fluorescence from these inverted $S_1$ states and unlocking new possibilities in OLED technology.[9–11]

This resurgence of interest in the inversion of the $S_1$ and $T_1$ excited state energies was primarily initiated by the computational chemistry community. In recent years, researchers have explored design strategies for the engineering of IST molecules with theoretical and computational methods.[12–16] However, the molecular structures resulting from these computational explorations have predominantly centered around nitrogen-doped phenalene-type structures, such as cyclazine or heptazine derivatives[1,2], along with related triangular[13,17] and hexagonal polycyclic aromatic hydrocarbons[18] with an interior filled by boron-nitride lattice.

Organic IST systems typically exhibit specific structural features and characteristics contributing to their unique electronic properties. While the structural aspects may vary, there are four common themes:

1. Conjugated π-systems: IST molecules represent a sub-class of polycyclic aromatic systems with electronic conjugation along the molecular edges.
2. Rigid, planar geometry: IST molecules typically possess rigid internal structures built from nitrogen, boron, or boron-nitride insulating lattices.
3. Electron-donating and electron-withdrawing groups: IST molecules often incorporate electron-donating and electron-withdrawing groups or atoms which allows the fine-tuning of their electronic properties.
4. Heavy-atom substituents: The incorporation of heavy atoms, such as sulfur, may increase the spin-orbit coupling in the molecule and may enhance the quantum yield of RISC.

While the concept of the organic IST chromophores appears promising, it is crucial to recognize an inherent challenge. Most IST compounds have thus far been found to exhibit minuscule oscillator strengths of the $S_1$-$S_0$ transition, resulting in low radiative decay rates.[12,14,15] The long radiative lifetime renders the fluorescence highly susceptible to competing non-radiative decay processes, compromising their emission efficiency.

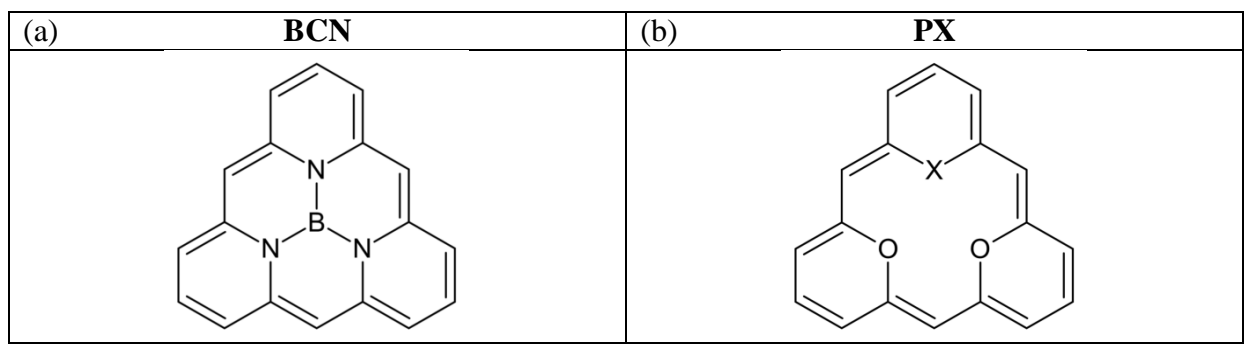

Chart 1. Triangular **BCN** (a) and boron-free (b) molecules considered in this work. X stands for O, NH, or S.

In this study, we computationally explored the optoelectronic properties of a novel category of the boron-free organic IST molecules, which are based on the triangle composed of the 2H-pyran units (**PX**) as illustrated in Chart 1. Our findings suggest the potential to modify the inherently negative $S_1$-$T_1$ energy gap, the wavelength of $S_1$-$S_0$ fluorescence, and its intensity through chemical alterations of the parent compounds. The results discussed herein represent a preliminary step towards constructing IST systems in the form of cyclic oligomers.

## 1. Computational methods

The ground-state equilibrium geometries of all compounds were in a first step optimized using density functional theory (DFT) employing the B3LYP functional,[19,20] augmented with Grimme's D3 dispersion correction.[21] The computation of the Hessian verified that the optimized stationary points represent energy minima. Vertical excitation energies of the lowest singlet and triplet states were computed with the second-order algebraic-diagrammatic construction (ADC(2)) method.[22–24]

To ensure consistency with the ADC(2) method employed for the excited-state calculations, the ground-state equilibrium geometries were re-optimized using the Møller-Plesset (MP2) method,[25] with the DFT equilibrium geometries serving as input. The ADC(2) method was also employed to determine the excited state ($S_1$ and $T_1$) equilibrium geometries. For all calculations, the correlation-consistent valence double-zeta cc-pVDZ basis set[26] was used. The calculations were performed with the Turbomole 7.3 program package.[27]

Benchmark calculations performed with the ADC(2) and with an approximate coupled cluster singles and doubles (CC2) methods indicate that the accuracies of both methods are very similar.[28] Previous studies have established that only methods explicitly including double excitations can accurately reproduce negative singlet-triplet energy gaps.[1,2,29–33] The ADC(2) method chosen for this study represents a judicious compromise between accuracy and computational cost.

## 2. Results and discussion
### 2.1. Vertical excitation energies

The fundamental structural concept underlying the triangular boron-free molecules examined in this study is illustrated in Chart 1. The defining principle of this new class of IST molecular systems investigated here entails the removal of the internal boron atom and the subsequent substitution of the remaining pyridine units with pyrans or its derivatives, while maintaining the electronic conjugation along the outer rim. This construction serves dual purposes: (i) it enables the creation of oligomeric molecular frameworks and (ii) it also introduces pyran moieties, thereby providing these molecules with unique electronic and structural properties.

The ground-state equilibrium geometries of the **BCN** molecule and the **PX** molecule with X = O (abbreviated **PO** in what follows), possess $D_{3h}$ symmetry, a feature that profoundly influences their electronic properties. The highest occupied molecular orbital (HOMO) and lowest unoccupied molecular orbital (LUMO) of both compounds are doubly degenerate. This orbital degeneracy gives rise to six valence excited states, three singlet states and three triplet states, with spatial symmetries $A_1'$, $A_2'$, and $E'$.

The vertical excitation energies of these lowest electronic states of the compounds shown in Chart 1 are listed in Table 1. Notably, in the symmetric systems represented by **BCN** and **PO** ($D_{3h}$ symmetry point group), the lowest excited singlet state emerges as a nondegenerate $^1A_2'(\pi\pi^*)$ state. It is interesting to note that the respective triplet ($^3A_2'$) state is energetically higher (negative singlet-triplet splitting) by 0.312 eV in **BCN**, and by 0.387 eV in **PO**. This is, however, not the lowest excited triplet state in both molecules because the singlet-triplet splitting of states with the $E'$ and $A_1'$ symmetry is positive and much larger than that of state with $A_2'$ symmetry. Thus, the lowest triplet state in **BCN** has $E'$ symmetry and $A_1'$ symmetry in **PO**. The $D_{3h}$ symmetry is reduced in asymmetrically substituted molecules, such as **PNH** (X = NH) and **PS** (X = S), which are of $C_{2v}$ and $C_s$ symmetry, respectively. This asymmetry removes the degeneracy of electronic states, as is seen in Table 1.

Table 1. Vertical excitation energies (in eV), oscillator strengths for the transition to the ground state (in parentheses) and the respective the $S_1$-$T_1$ energy gap ($\Delta_{ST}$) of the structures presented in Chart 1 computed at the ADC(2) level of theory.

| **BCN**($D_{3h}$) | | **PO**($D_{3h}$) | | **PNH**($C_{2v}$) | | **PS**($C_s$) | |
|---|---|---|---|---|---|---|---|
| State | E/eV | State | E/eV | State | E/eV | State | E/eV |
| triplet states | | | | | | | |
| $^3E'$ | **1.914** | $^3A_1'$ | **1.911** | $^3B_1$ | **1.734** | $^3A''$ | **1.550** |
| $^3E'$ | 1.914 | $^3E'$ | 1.937 | $^3A_1$ | 1.941 | $^3A'$ | 1.552 |
| $^3A_1'$ | 1.934 | $^3E'$ | 1.937 | $^3A_1$ | 2.111 | $^3A'$ | 1.832 |
| $^3A_2'$ | 1.943 | $^3A_2'$ | 1.984 | $^3B_1$ | 2.371 | $^3A''$ | 2.137 |
| singlet states | | | | | | | |
| $^1A_2'$ | **1.631** (0.0) | $^1A_2'$ | **1.597** (0.0) | $^1B_1$ | **1.620**(0.037) | $^1A''$ | **1.446**(0.023) |
| $^1A_1'$ | 2.253 (0.0) | $^1A_1'$ | 2.254 (0.0) | $^1A_1$ | 2.398(0.005) | $^1A'$ | 2.006(0.017) |
| $^1E'$ | 2.370 (0.293) | $^1E'$ | 2.572 (0.468) | $^1A_1$ | 2.659(0.509) | $^1A'$ | 2.424(0.400) |
| $^1E'$ | 2.370 (0.293) | $^1E'$ | 2.572 (0.468) | $^1B_1$ | 2.699(0.259) | $^1A''$ | 2.549(0.458) |
| $\Delta_{ST}$ | **-0.283** | $\Delta_{ST}$ | **-0.314** | $\Delta_{ST}$ | **-0.114** | $\Delta_{ST}$ | **-0.104** |

Remarkably, all the species included in Table 1 are IST systems, that is, the $S_1$-$T_1$ energy gap, defined as $\Delta_{ST} = E_S - E_T$, is negative, where $E_S$ and $E_T$ are the lowest singlet and lowest triplet state energies. A negative $\Delta_{ST}$ signifies that the vertical excitation energy of the $T_1$ state exceeds that of the $S_1$ state. In the symmetric **PO** and **BCN** molecules, the $S_0$-$S_1$ transition is symmetry forbidden (f = 0). Breaking the trigonal symmetry axis by substituting one of the oxygen atoms of **PO** with a NH group (in **PNH**), or with sulfur (in **PS**), induces a nonzero dipole moment for this transition. The reduction of the symmetry is also accompanied by a decrease in the magnitude of the negative singlet-triplet energy gap from -0.283 for **BCN** to -0.104 for **PS**.

To gain deeper insights into this phenomenon, Fig. 1 depicts the electron density of molecular orbitals involved in the lowest-energy electronic transitions. The electron densities presented in Fig. 1 were computed assuming equal occupation of both components of the degenerate E' HOMO and LUMO orbitals in **BCN** and **PO**. Asymmetric substitutions to the **PX** system remove orbital degeneracy, but still, the $S_1$ state in both (**PNH** and **PS**) molecules, contains comparable contributions from HOMO/LUMO and HOMO-1/LUMO+1 transitions. Thus for the sake of transparency, the assumption of equal occupation of HOMO and HOMO-1 as well as LUMO and LUMO+1 orbitals were used in computation of electron density involved in this electronic transition.

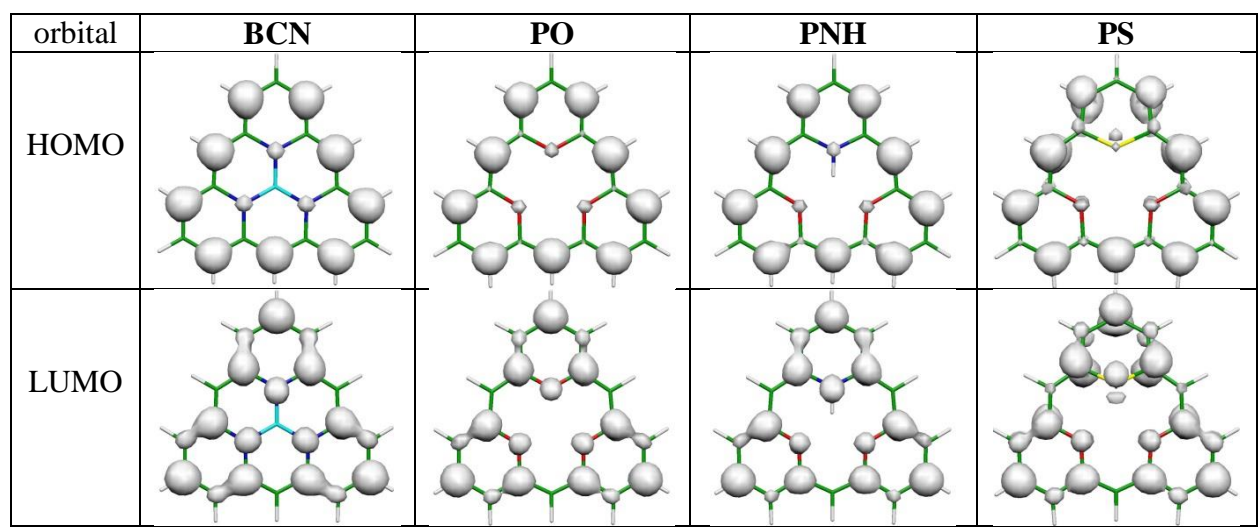

| orbital | **BCN** | **PO** | **PNH** | **PS** |
|---|---|---|---|---|
| HOMO | | | | |
| LUMO | | | | |

Fig. 1. Electron density of (near)degenerate HOMO and LUMO orbitals of the compounds presented in Chart 1.

Fig. 1 reveals that the distribution of the electron densities of HOMO and LUMO display a characteristic pattern which has been found to be typical for IST systems.[18] For the HOMO and, to a lesser extent, for the LUMO, the electronic charge distribution is expelled from the interior of the molecular framework. Instead, it is largely localized on alternating atoms along the rim. This distribution minimizes the exchange integral, which results in a small $S_1$-$T_1$ splitting. However, it also reduces the transition dipole moment and thus the oscillator strength of the $S_0$-$S_1$ transition. Chemical substitutions that reduce the symmetry, such as those in **PNH** and **PS**, increase the overlap between electron densities. As mentioned earlier, this affects both the $S_1$-$T_1$ energy gap and the oscillator strength of the $S_0$-$S_1$ transition.

### 2.2. Effect of additional symmetry reduction by chemical modification

While a negative $\Delta_{ST}$ represents an enticing prospect for applications in OLED devices, the low fluorescence intensity of the $S_1$ state, as mentioned above, poses a significant drawback. To delve deeper into the impact of symmetry reduction on the photophysical properties of IST molecules, an overview of the computed spectroscopic parameters for four specifically chosen chemically modified **PX** systems is provided in Table 2. Additionally, an extensive selection of chemically modified **PX** systems is presented in the Electronic Supplementary Information (ESI), offering a broader exploration of the effects of symmetry alteration on their energy level structure.

Table 2. Vertical excitation energies (in eV) of the lowest excited singlet and triplet states, oscillator strengths for the $S_0 \rightarrow S_1$ transition (in parentheses), and the respective $S_1$-$T_1$ energy gap ($\Delta_{ST}$) of the selected **PX** molecules, computed at the ADC(2) level of theory.

| State | **PO-4N** 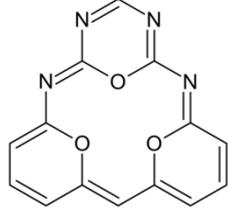 | **PNH-4N** 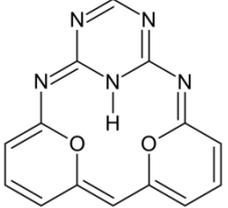 | **PS-2N** 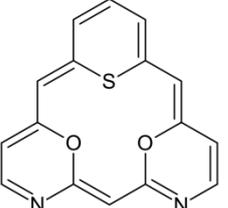 | **PNH-2NO$_2$** 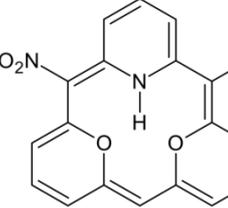 |
|---|---|---|---|---|
| $T_1$ | 1.981 | 1.796 | 1.623 | 1.796 |
| $S_1$ | 1.953(0.072) | 1.761(0.086) | 1.519(0.039) | 1.792(0.132) |
| $\Delta_{ST}$ | -0.028 | -0.035 | -0.104 | -0.004 |

Table 2 demonstrates that appropriate chemical modifications to the molecular structure can notably enhance the oscillator strength of the $S_0$-$S_1$ transition. However, this effect correlates strongly with a reduction in the magnitude of the negative singlet-triplet splitting: the greater the oscillator strength of the transition, the smaller the negative singlet-triplet energy gap. Bearing in mind that the ADC(2) method tends to overestimate this effect,[34,35] it can be concluded that the molecules listed in the table represent TADF systems with exceptionally small positive ST splitting.

The correlation between the oscillator strength and the singlet-triplet energy gap is revealed by the correlation diagram presented in Fig. 2, which showcases three families of **PX** (X = O,

NH, S) systems featuring nitrogen substitutions at the molecular rim. Apart from the systems with symmetry-forbidden $S_0$-$S_1$ transitions (f = 0), a clear overall correlation between f and $\Delta_{ST}$ is discernible. On the other hand, the scattering of the symbols in the diagram reveals nuanced possibilities of manipulating $\Delta_{ST}$ and f by molecular symmetry reduction *via* CH/N substitutions at the rim. This variability highlights the flexibility inherent in designing **PX** systems tailored to exhibit specific desired properties, thereby opening up exciting possibilities for manipulating their photophysical functionality.

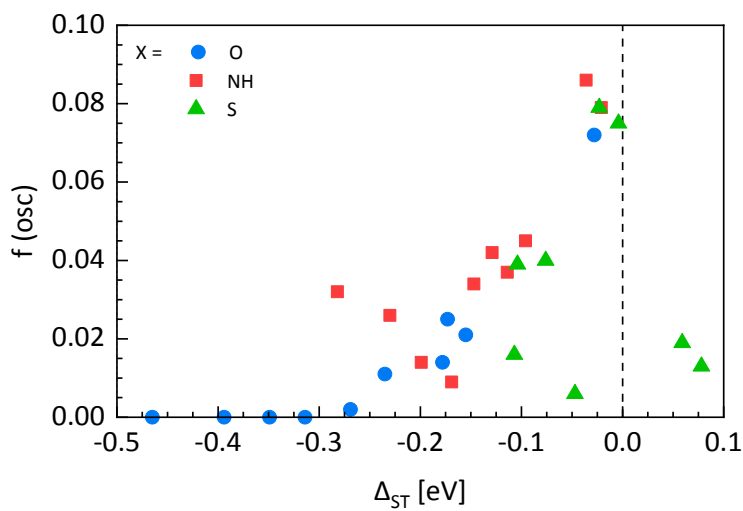

Fig. 2. Correlation between the $S_1$-$T_1$ energy gap and the oscillator strength of the $S_0 \rightarrow S_1$ transition with different carbon/nitrogen replacements at the molecular rim (**PO** - blue circles, **PNH** - red squares, **PS** - green triangles), computed at the ADC(2) level of theory. (see Tables S1, S3, and S5 in the SI)

Another intriguing aspect of the computed photophysical properties of the **PX** molecular systems is the correlation between the energy of the $S_0$-$S_1$ transition and the number of nitrogen atoms present in the molecular rim. Fig. 3 shows that an increased number of nitrogen atoms increases the energy of the $S_0$-$S_1$ transition and enhances the magnitude of the singlet-triplet inversion (for clarity, only nitrogen replacements maintaining the triple symmetry axis of the **PX** system are included in Fig. 3). This observation underscores the intricate relationship between the molecular composition and photophysical properties which can be exploited for the tailored design of OLED chromophores. The impact of the asymmetric CH/N substitutions at the molecular rim on the energy of the $S_0$-$S_1$ transition and its oscillator strength is additionally documented in the ESI.

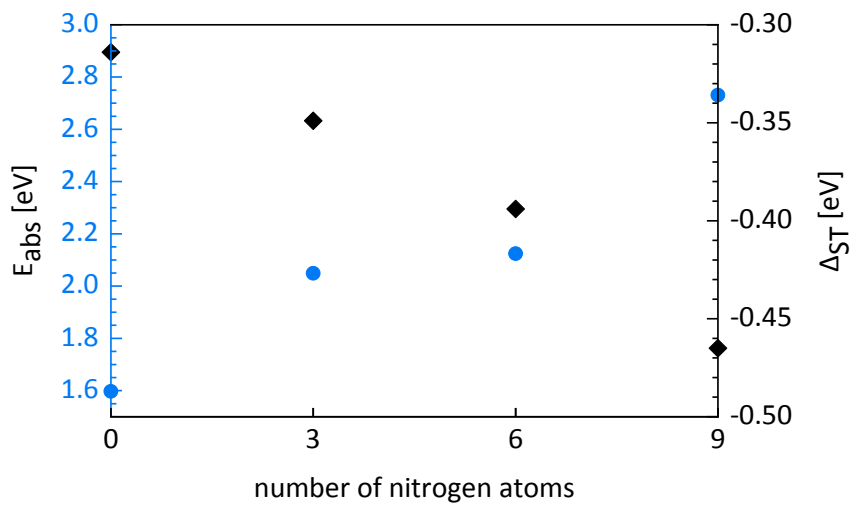

Fig. 3. Correlation between the energy of the $S_0$-$S_1$ transition (left axis, blue circles) and the $S_1$-$T_1$ energy gap (right axis, black rhombs), plotted against the number of nitrogen atoms present in the molecular rim (molecules **PO-N**, **PO-3Na**, **PO-6N**, **PO-9N** from Table S1 of the SI), obtained with the ADC(2) method.

### 3.3 Radiative emission properties

Spectral characteristics discussed thus far were determined for the ground-state equilibrium geometries of the **PX** systems. However, for the characterization of radiative emissions (fluorescence, delayed fluorescence, and phosphorescence), it is crucial to consider the photophysical properties at the equilibrium geometries of the excited states ($S_1$ and $T_1$). The vertical excitation energies in Table 1 show that in the molecules with three-fold symmetry axes (**BCN** and **PO**), the lowest excited singlet state $S_1(A_2')$ is well separated from higher excited singlet states. The two lowest excited triplet states ($^3A_1'$ and $^3E'$), on the other hand, are nearly degenerate. While geometry optimization of the non-degenerate singlet state maintains $D_{3h}$ symmetry, the degenerate (E') state may exhibit molecular symmetry breaking due to the Jahn-Teller (JT) effect.

The ADC(2) method, as currently implemented in TURBOMOLE package, is limited to dealing with Abelian symmetry point groups. The highest Abelian subgroup of $D_{3h}$ is $C_{2v}$. Under $C_{2v}$ symmetry, the $A_2'$ symmetry representation becomes $B_1$, and the two degenerate components of the E' representation transform as $A_1$ and $B_1$, respectively. While the **BCN** molecule maintains $D_{3h}$ symmetry when optimizing the geometry of the $S_1(B_1)$ state, this symmetry is lost when the geometries of the $^3B_1$ and $^3A_1$ states are optimized. The Hessian computed at these stationary points indicates that the optimized $^3A_1$ state represents a local minimum, while the optimized $^3B_1$ state is a first-order saddle point of the JT-deformed two-dimensional PES of the $^3E'$ state. Its relative energy (0.24 eV) with respect to the minimum of the $^3A_1$ state represents the energy barrier for the so-called pseudo-rotation on the PES of the $T_1$ state.

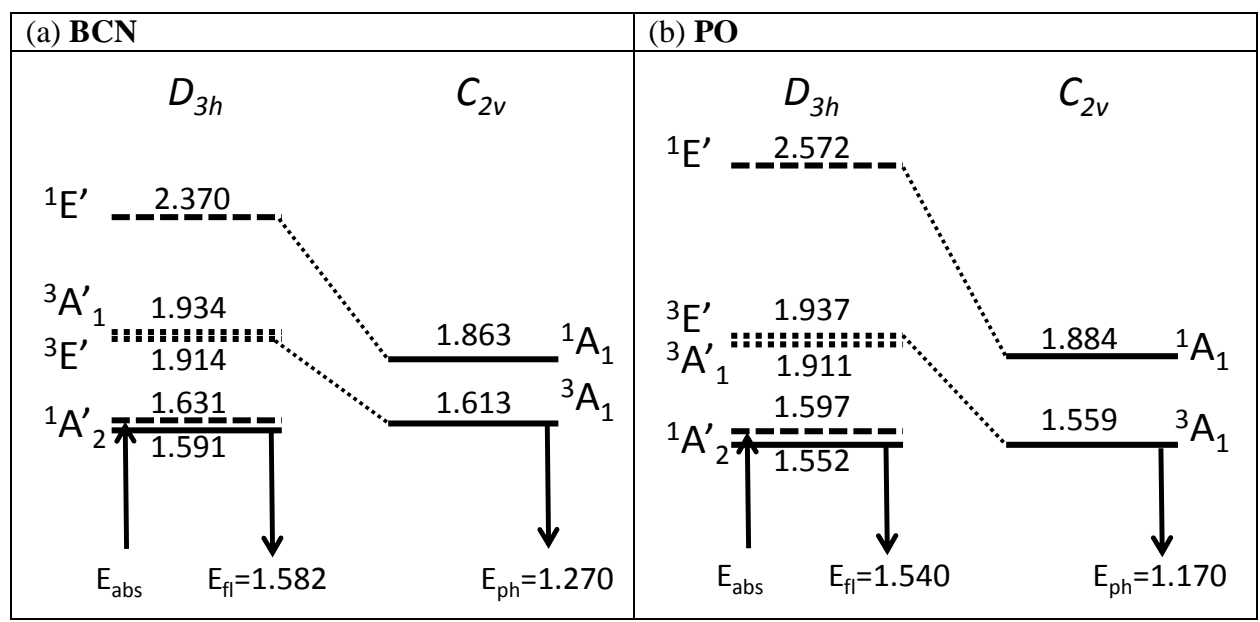

Fig. 4. Energy-level schemes of the **BCN** (a) and **PO** (b) molecules determined at the ADC(2) level of theory. Solid lines denote the optimized energy levels of the respective electronically excited states, while dashed (dotted) lines denote the vertical energy of the singlet (triplet) states computed at the equilibrium geometry of the electronic ground state. Up and down arrows denote vertical absorption and emission, respectively. Numbers denote energies in electronvolts.

A qualitatively similar energy-level scheme is obtained for the **PO** molecule. Both dschemes are shown in Fig. 4. It is evident that the non-degenerate $S_1(A_2')$ state ($^1B_1$ symmetry within the $C_{2v}$ point group) of both molecules exhibits rigidity, that is, the energy relaxation upon geometry optimization is small (0.040 eV and 0.045 eV, respectively) and does not break molecular symmetry. For the degenerate triplet state $^3E'$, on the contrary, the energy relaxation upon geometry optimization is much more pronounced (0.301 eV and 0.378 eV, respectively). A similarly pronounced relaxation of the energy is also observed in the $^1E'$ state. In both E' states, this effect is attributed to JT-induced geometric instability.

Several interesting conclusions emerge from Fig. 4. Particularly noteworthy is the singlet-triplet inversion measured by the vertical energy difference computed at the ground-state equilibrium geometry, which is remarkably large (-0.283 eV for **BCN** and -0.314 eV for **PO**).

However, the corresponding difference between the adiabatic (geometry-optimized) energies of singlet and triplet states, which represents the difference in the 0-0 transition energy, is much smaller (-0.022 eV and -0.007 eV, respectively). Furthermore, the difference between the vertical fluorescence ($S_1$-$S_0$) and phosphorescence ($T_1$-$S_0$) energies, representing the difference of the peak maxima of the fluorescence ($E_{fl}$) and phosphorescence ($E_{ph}$) spectra, becomes positive (0.312 eV and 0.370 eV, respectively).

A comparison of the level schemes of both molecules clearly shows that the removal of the central boron atom from **BCN** and the replacement of the remaining nitrogen atoms by oxygen atoms (**PO**) have a minor effect on their photophysics. Despite the removal of the central skeleton atom, the **PO** molecule exhibits rigidity which is similar to **BCN**. The inversion of the vertical singlet and triplet states is strongly tied to the electronic conjugation along the molecular rim.

Replacing one of the oxygen atoms in **PO** with the NH group conserves the number of electrons in the system, but reduces the molecular symmetry to $C_{2v}$ or lower. The relevant energy-level schemes for **PNH-4N** (Table 2) and **PNH-6N** (Table S3 in the SI) where six nitrogen atoms are symmetrically distributed at the corners of pyran units (see ESI), are shown in Fig. 5. It is evident that geometry optimization of the lowest singlet and triplet states of **PNH-4N** (both having the $B_1$ symmetry in the $C_{2v}$ point group) does not further lower the molecular symmetry and stabilizes both states by nearly the same amount of energy (0.076 eV for the $^1B_1$ state and 0.069 for the $^3B_1$ state). The adiabatic energies of these states are inverted by merely -0.042 eV. The peak energy of phosphorescence is predicted to be only 0.03 eV lower than the peak energy of fluorescence.

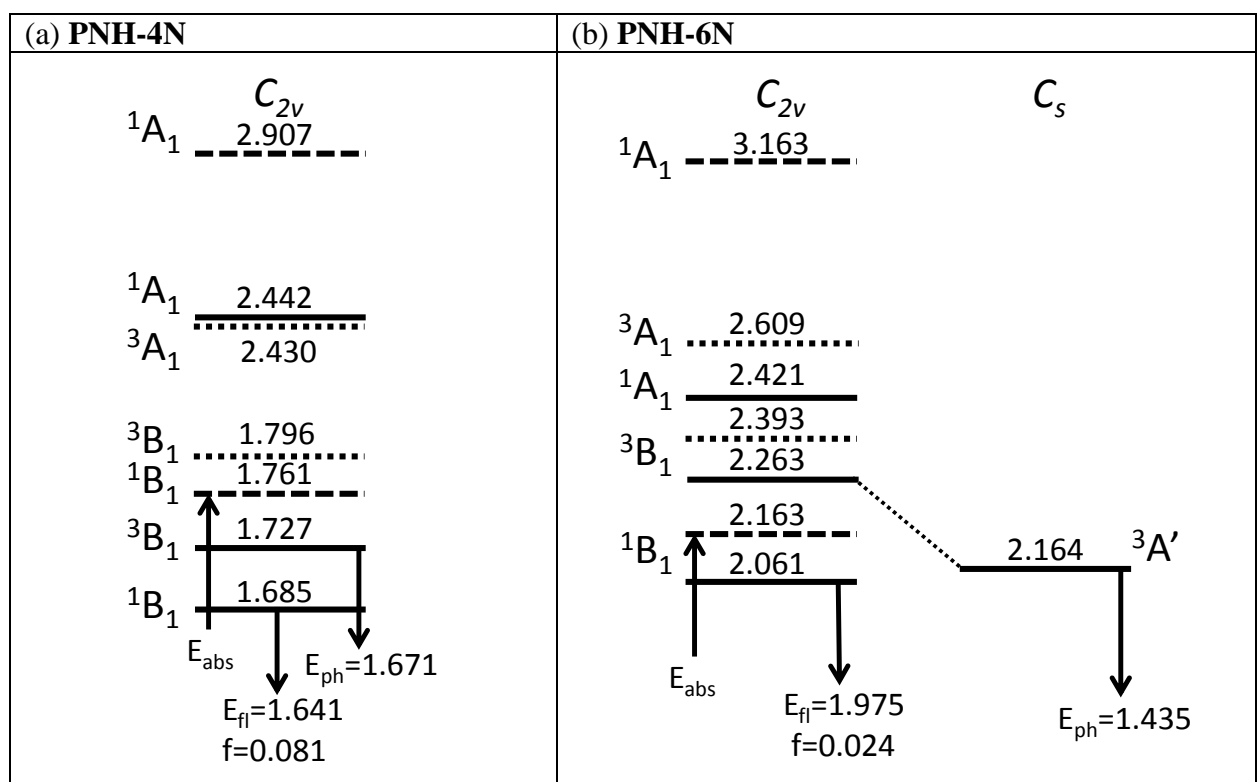

Fig. 5. Energy-level schemes of **PNH-4N** (a) and **PNH-6N** (b) molecules determined at the ADC(2) level of theory. Solid lines denote the optimized energy levels of the respective electronically excited states, and dashed (dotted) lines denote the vertical energy of the singlet (triplet) states computed at the equilibrium geometry of the ground state. Up and down arrows denote vertical absorption and emission, respectively. Numbers denote energy in electronvolts.

A symmetric distribution of six nitrogen atoms along the molecular rim of the **PNH** molecule (**PNH-6N**) maintains $C_{2v}$ symmetry, but notably decreases the energy gap between the $^3B_1$ and $^3A_1$ states (Fig. 5b). While unconstrained geometry optimization of the lowest excited singlet state conserves $C_{2v}$ symmetry, such optimization of the lowest triplet state results in symmetry lowering to $C_s$ (only the molecular plane is conserved). The magnitude of the negative $S_1$-$T_1$ vertical energy gap (-0.230 eV) decreases to -0.103 eV for the adiabatic energies. The difference of the emission maxima from these states becomes positive (0.540 eV). Inspection of the equilibrium geometry of the $^3A'$ state (Table S6 of the SI) reveals that, apart from the JT distortion, a significant amount of Kekule-like single-double CC bond alternation occurs.

## 3. Summary

Pyran-substituted triangular aromatic hydrocarbons represent a new category of molecular systems that can be tailored to exhibit robustly negative singlet-triplet energy gaps. The results of this computational study confirm that previously discussed triangular carbon nitrides and triangular boron carbon nitrides are not the only stable organic molecules capable of displaying $S_1$-$T_1$ inversion.

A common feature of these IST systems is the localization of electronic conjugation at the molecular rim rather than in the interior. This effect can be achieved by either filling the interior of an aromatic arene or azaarene molecule with an insulating motif (such as boron nitride) or by removing the central atom altogether, as in the systems examined in this study. Strictly speaking, the triangular molecules investigated in this work can be considered as cyclic oligomers composed of the pyran units and its derivatives. This finding paves a new path for constructing IST systems utilizing organic molecular units.

There is a consensus in the computational literature that the vertical energy gap between the $S_1$ and $T_1$ states, computed at the equilibrium geometry of the ground state, defines the ST gap. However, from the perspective of photophysical processes occurring between these states, such as reverse intersystem crossing (RISC), the adiabatic energy of these states becomes significant. This refers to the energy of the so-called 0-0 line in the fluorescence and phosphorescence spectra. The results of this study clearly demonstrate that, in many cases, these are two distinct quantities.

**Acknowledgements**

The Authors would like to thank Prof. Wolfgang Domcke for the fruitful discussions upon preparation of the manuscript. This research was funded by National Science Centre of Poland, grant number: 2020/39/B/ST4/01723. We gratefully acknowledge Polish high-performance computing infrastructure PLGrid (HPC Center: ACK Cyfronet AGH) for providing computer facilities and support within computational grant no. PLG/2024/017058.

**Conflicts of interest**

There are no conflicts to declare.

# Electronic Supplementary Information (ESI)

## Computational Design of Boron-Free Triangular Molecules with Inverted Singlet-Triplet Energy Gap

M.W. Duszka, M.F. Rode, and A.L. Sobolewski

Institute of Physics, Polish Academy of Sciences, Warsaw, Poland

**Table S1**. Vertical absorption energy of the lowest singlet ($E(S_1)$), and lowest triplet ($E(T_1)$) states (spatial symmetry labels in parenthesis), energy gap between the lowest singlet and lowest triplet states ($\Delta_{ST} = E(S_1)-E(T_1)$, in eV), and oscillator strength (f), determined with the ADC(2)/cc-pVDZ method at the ground-state equilibrium geometry optimized at the MP2/cc-pVDZ theory level of **PO** and its derivatives with nitrogen substitutions within the external rim.

| abbreviation | molecule | $E(S_1)$ $E(T_1)$ $\Delta_{ST}$ f | abbreviation | molecule | $E(S_1)$ $E(T_1)$ $\Delta_{ST}$ f |
|---|---|---|---|---|---|
| **PO-N** | 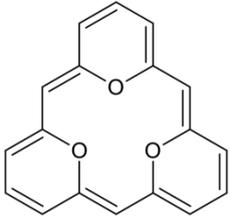 | 1.597 ($^1B_1$) 1.911 ($^3A_1$) -0.314 0.000 | **PO-3N** | 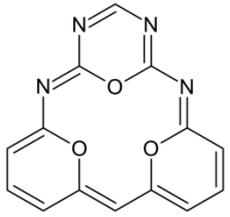 | 2.049 ($^1B_1$) 2.398 ($^3B_1$) -0.349 0.000 |
| **PO-2Na** | 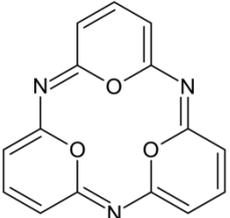 | 1.689 ($^1B_1$) 1.862 ($^3B_1$) -0.173 0.025 | **PO-4Na** | 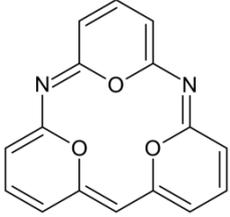 | 1.953 ($^1B_1$) 1.981 ($^3B_1$) -0.028 0.072 |
| **PO-2Nb** | 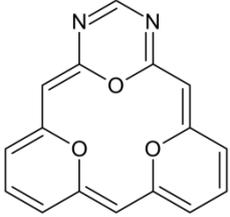 | 1.873 ($^1B_1$) 2.051 ($^3B_1$) -0.178 0.014 | **PO-4Nb** | 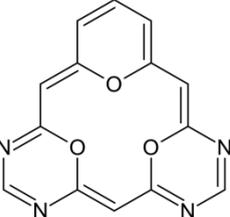 | 1.888 ($^1B_1$) 2.043 ($^3B_1$) -0.155 0.021 |

| abbreviation | molecule | E(S$_1$)<br>E(T$_1$)<br>$\Delta_{ST}$<br>f | abbreviation | molecule | E(S$_1$)<br>E(T$_1$)<br>$\Delta_{ST}$<br>f |
|---|---|---|---|---|---|
| **PO-2Nc** | 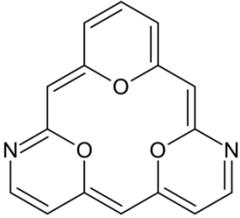 | 1.720 ($^1$B$_1$)<br>1.989 ($^3$B$_1$)<br>-0.269<br>0.002 | **PO-6N** | 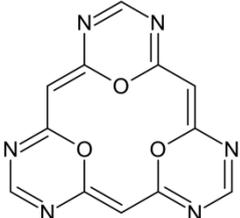 | 2.124 ($^1$B$_1$)<br>2.518 ($^3$B$_1$)<br>-0.394<br>0.000 |
| **PO-2Nd** | 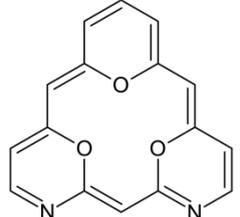 | 1.732 ($^1$B$_1$)<br>1.976 ($^3$B$_1$)<br>-0.235<br>0.011 | **PO-9N** | 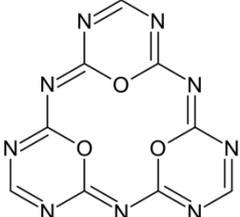 | 2.731 ($^1$B$_1$)<br>3.196 ($^3$B$_1$)<br>-0.465<br>0.000 |

**Table S2**. Vertical absorption energy of the lowest singlet (E(S$_1$)), and lowest triplet (E(T$_1$)) states (spatial symmetry labels in parenthesis), energy gap between the lowest singlet and lowest triplet states ($\Delta_{ST}$ = E(S$_1$)-E(T$_1$), in eV), and oscillator strength (f), determined with the ADC(2)/cc-pVDZ method at the ground-state equilibrium geometry optimized at the MP2/cc-pVDZ theory level of **PO** derivatives with amino and nitro substitutions.

| abbreviation | molecule | E(S$_1$)<br>E(T$_1$)<br>$\Delta_{ST}$<br>f | abbreviation | molecule | E(S$_1$)<br>E(T$_1$)<br>$\Delta_{ST}$<br>f |
|---|---|---|---|---|---|
| **PO-NO$_2$** | 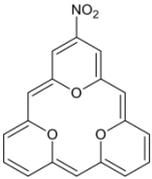 | 1.499 ($^1$B$_1$)<br>1.548 ($^3$B$_1$)<br>-0.049<br>0.010 | **PO-2NO$_2$b** | 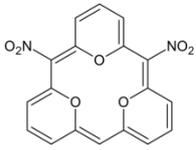 | 1.870 ($^1$B$_1$)<br>1.969 ($^3$B$_1$)<br>-0.099<br>0.017 |
| **PO-NH$_2$** | 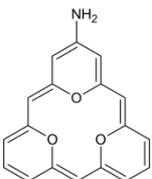 | 1.669 ($^1$A")<br>1.893 ($^3$A")<br>-0.224<br>0.014 | **PO-2NH$_2$a** | 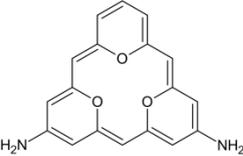 | 1.772 ($^1$A")<br>1.965 ($^3$A")<br>-0.193<br>0.017 |
| **PO-2NO$_2$a** | 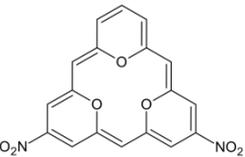 | 1.434 ($^1$B$_1$)<br>1.442 ($^3$A$_1$)<br>-0.008<br>0.007 | **PO-2NH$_2$b** | 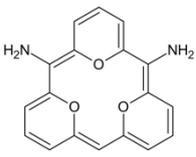 | 1.763 ($^1$A)<br>1.790 ($^3$A)<br>-0.027<br>0.001 |

**Table S3**. Vertical absorption energy of the lowest singlet (E(S$_1$)), and lowest triplet (E(T$_1$)) states (spatial symmetry labels in parenthesis), energy gap between the lowest singlet and lowest triplet states ($\Delta_{ST}$ = E(S$_1$)-E(T$_1$), in eV), and oscillator strength (f), determined with the ADC(2)/cc-pVDZ method at the ground-state equilibrium geometry optimized at the MP2/cc-pVDZ theory level of **PNH** derivatives with nitrogen substitutions to the external rim.

| abbreviation | molecule | E(S$_1$)<br>E(T$_1$)<br>$\Delta_{ST}$<br>f | abbreviation | molecule | E(S$_1$)<br>E(T$_1$)<br>$\Delta_{ST}$<br>f |
|---|---|---|---|---|---|
| **PNH** | 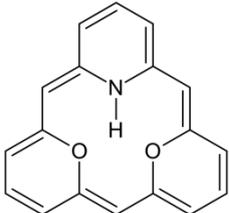 | 1.620 ($^1$B$_1$)<br>1.734 ($^3$B$_1$)<br>-0.114<br>0.037 | **PNH-3N** | 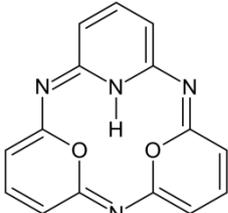 | 2.096 ($^1$B$_1$)<br>2.192 ($^3$B$_1$)<br>-0.096<br>0.045 |
| **PNH-2Na** | 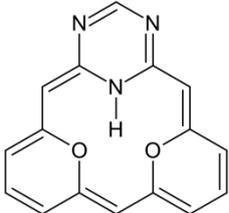 | 1.477 ($^1$B$_1$)<br>1.606 ($^3$B$_1$)<br>-0.129<br>0.042 | **PNH-4Na** | 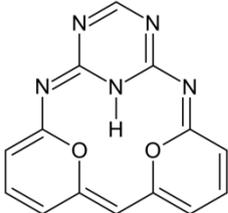 | 1.761 ($^1$B$_1$)<br>1.796 ($^3$B$_1$)<br>-0.035<br>0.086 |
| **PNH-2Nb** | 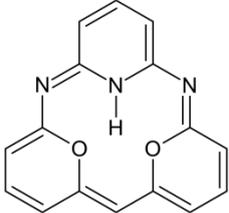 | 1.869 ($^1$B$_1$)<br>1.890 ($^3$B$_1$)<br>-0.021<br>0.079 | **PNH-4Nb** | 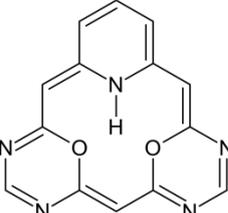 | 2.218 ($^1$B$_1$)<br>2.387 ($^3$A$_1$)<br>-0.169<br>0.009 |
| **PNH-2Nc** | 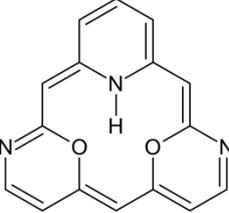 | 1.869 ($^1$B$_1$)<br>2.016 ($^3$B$_1$)<br>-0.147<br>0.034 | **PNH-6N** | 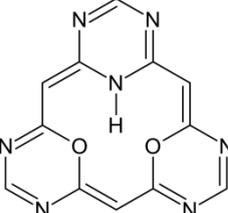 | 2.163 ($^1$B$_1$)<br>2.393 ($^3$B$_1$)<br>-0.230<br>0.026 |
| **PNH-2Nd** | 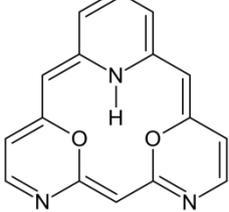 | 1.939 ($^1$B$_1$)<br>2.138 ($^3$B$_1$)<br>-0.199<br>0.014 | **PNH-9N** | 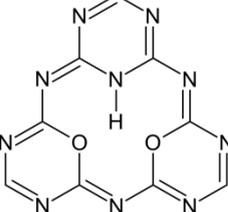 | 2.778 ($^1$B$_1$)<br>3.060 ($^3$B$_1$)<br>-0.282<br>0.032 |

**Table S4**. Vertical absorption energy of the lowest singlet (E(S$_1$)), and lowest triplet (E(T$_1$)) states (spatial symmetry labels in parenthesis), energy gap between the lowest singlet and lowest triplet states ($\Delta_{ST}$ = E(S$_1$)-E(T$_1$), in eV), and oscillator strength (f), determined with the ADC(2)/cc-pVDZ method at the ground-state equilibrium geometry optimized at the MP2/cc-pVDZ theory level of **PNH** derivatives with amino and nitro substitutions.

| abbreviation | molecule | E(S$_1$) E(T$_1$) $\Delta_{ST}$ f | abbreviation | molecule | E(S$_1$) E(T$_1$) $\Delta_{ST}$ f |
|---|---|---|---|---|---|
| **PHN-NO$_2$** | 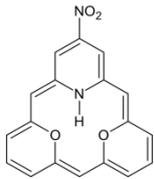 | 1.533 ($^1$B$_1$) 1.731 ($^3$B$_1$) -0.198 0.012 | **PNH-2NO$_2$b** | 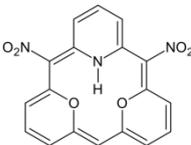 | 1.792 ($^1$B$_1$) 1.796 ($^3$B$_1$) -0.004 0.132 |
| **PNH-NH$_2$** | 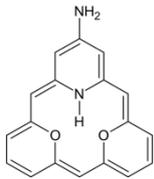 | 1.634 ($^1$A") 1.703 ($^3$A") -0.069 0.058 | **PNH-2NH$_2$a** | 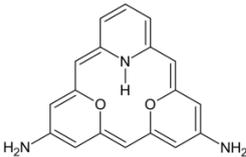 | 2.705 ($^1$A) 2.249 ($^3$A) 0.456 0.000 |
| **PNH-2NO$_2$a** | 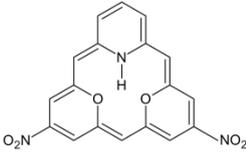 | 1.432 ($^1$B$_1$) 1.348 ($^3$B$_1$) 0.084 0.064 | **PNH-2NH$_2$b** | 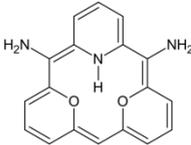 | 2.328 ($^1$A$_1$) 1.827 ($^3$A$_1$) 0.501 0.026 |

**Table S5**. Vertical absorption energy of the lowest singlet (E(S$_1$)), and lowest triplet (E(T$_1$)) states (spatial symmetry labels in parenthesis), energy gap between the lowest singlet and lowest triplet states ($\Delta_{ST}$ = E(S$_1$)-E(T$_1$), in eV), and oscillator strength (f), determined with the ADC(2)/cc-pVDZ method at the ground-state equilibrium geometry optimized at the MP2/cc-pVDZ theory level of **PS** derivatives with nitrogen substitutions to the external rim.

| abbreviation | molecule | E(S$_1$)<br>E(T$_1$)<br>$\Delta_{ST}$<br>f | abbreviation | molecule | E(S$_1$)<br>E(T$_1$)<br>$\Delta_{ST}$<br>f |
|---|---|---|---|---|---|
| **PS-N** | 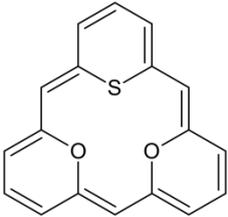 | 1.446 ($^1$A")<br>1.550 ($^3$A")<br>-0.104<br>0.023 | **PS-4Nb** | 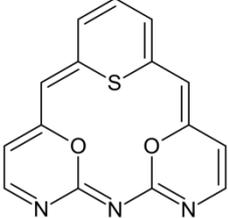 | 1.656 ($^1$A")<br>1.732 ($^3$A")<br>-0.076<br>0.040 |
| **PS-2N** | 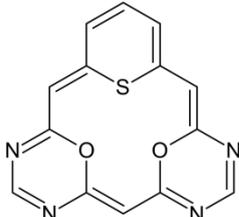 | 1.519 ($^1$A")<br>1.623 ($^3$A")<br>-0.104<br>0.039 | **PS-4Nc** | 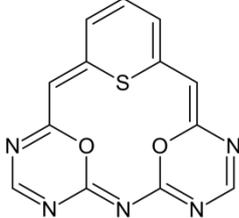 | 1.908 ($^1$A")<br>1.955 ($^3$A")<br>-0.047<br>0.006 |
| **PS-3Nb** | 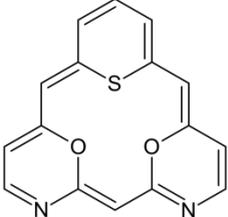 | 1.600 ($^1$A")<br>1.623 ($^3$A")<br>-0.023<br>0.079 | **PS-5N** | 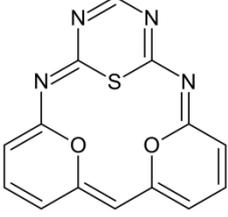 | 1.726 ($^1$A")<br>1.730 ($^3$A")<br>-0.004<br>0.075 |
| **PS-4Na** | 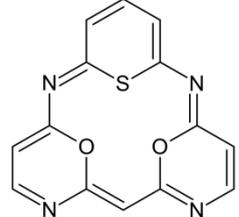 | 1.948 ($^1$A")<br>1.870 ($^3$A")<br>0.078<br>0.013 | **P2S** | 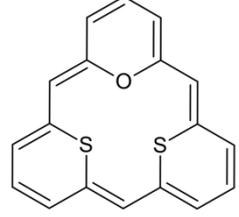 | 1.485 ($^1$A)<br>1.422 ($^3$A)<br>0.059<br>0.019 |

**Table S6.** Lengths of the external CC/CN bonds in the optimized molecular geometries of the ground ($S_0$), and of the lowest excited singlet ($S_1$) and triplet ($T_1$) states, their adiabatic energy ($E_{0-0}$), energy of the $S_1 \rightarrow S_0$ or $T_1 \rightarrow S_0$ vertical emission fluorescence and phosphorescence ($E_{em}$), the adiabatic energy difference between the lowest excited singlet and triplet states ($\Delta_{ST} = S_1-T_1$), and oscillator strength ($f_{em}$) from the $S_1$ state of selected PX molecules. Molecular symmetry in a given state is in parentheses, and energies are in electronvolts.

| | **BCN** | | | |
|---|---|---|---|---|
| | $S_0(D_{3h})$ | $S_1(D_{3h})$ | $T_1(C_{2v})$ | $\Delta_{ST}$ |
| | (structure) | (structure) | (structure) | |
| $E_{0-0}$ | - | 1.591 | 1.613 | -0.022 |
| $E_{em}$ | - | 1.582 | 1.270 | 0.312 |
| $f_{em}$ | - | 0.0 | - | |
| | **PO** | | | |
| | $S_0(D_{3h})$ | $S_1(D_{3h})$ | $T_1(C_{2v})$ | $\Delta_{ST}$ |
| | (structure) | (structure) | (structure) | |
| $E_{0-0}$ | - | 1.552 | 1.559 | -0.007 |
| $E_{em}$ | - | 1.540 | 1.170 | 0.370 |
| $f_{em}$ | - | 0.0 | - | |
| | **PNH-4N** | | | |
| | $S_0(C_{2v})$ | $S_1(C_{2v})$ | $T_1(C_{2v})$ | $\Delta_{ST}$ |
| | (structure) | (structure) | (structure) | |
| $E_{0-0}$ | - | 1.685 | 1.727 | -0.042 |
| $E_{em}$ | - | 1.641 | 1.671 | -0.030 |
| $f_{em}$ | - | 0.081 | - | |

| | **PNH-6N** | | | |
|---|---|---|---|---|
| | $T_1(C_{2v})$ | $S_1(C_{2v})$ | $S_1(C_s)$ | $\Delta_{ST}$ |
| | | | | |
| $E_{0-0}$ | - | 2.061 | 2.164 | -0.103 |
| $E_{em}$ | - | 1.975 | 1.435 | 0.540 |
| $f_{em}$ | - | 0.024 | - | |